\documentclass[twocolumn,pra,showpacs,superscriptaddress]{revtex4-1}
\usepackage{graphicx}
\usepackage{subfigure}
\usepackage{amsmath}
\usepackage{amsfonts,amssymb}
\usepackage{bm}
\usepackage{float}
\usepackage{color}
\usepackage{sidecap}
\allowdisplaybreaks
\usepackage{soul}
\setstcolor{red}
\usepackage{ulem}
\usepackage{hyperref}
\hypersetup{pdfpagemode=FullScreen,colorlinks=true,breaklinks,urlcolor=blue,linkcolor=blue,citecolor=blue}

\begin{document}

\title{Anderson localization in the Non-Hermitian Aubry-Andr\'e-Harper model with physical gain and loss}

\author{Qi-Bo Zeng}
\email{zqb15@mails.tsinghua.edu.cn}
\affiliation{Department of Physics and State Key Laboratory of Low-Dimensional Quantum Physics, Tsinghua University, Beijing 100084, China}

\author{Shu Chen}
\affiliation{Beijing National Laboratory for Condensed Matter Physics, Institute of Physics,
Chinese Academy of Sciences, Beijing 100190, China}
\affiliation{Collaborative Innovation Center of Quantum Matter, Beijing 100084, China}

\author{Rong L\"u}
\email{rlu@tsinghua.edu.cn}
\affiliation{Department of Physics and State Key Laboratory of Low-Dimensional Quantum Physics, Tsinghua University, Beijing 100084, China}
\affiliation{Collaborative Innovation Center of Quantum Matter, Beijing 100084, China}

\begin{abstract}
We investigate the Anderson localization in non-Hermitian Aubry-Andr\'e-Harper (AAH) models with imaginary potentials added to lattice sites to represent the physical gain and loss during the interacting processes between the system and environment. By checking the mean inverse participation ratio (MIPR) of the system, we find that different configurations of physical gain and loss have very different impacts on the localization phase transition in the system. In the case with balanced physical gain and loss added in an alternate way to the lattice sites, the critical region (in the case with p-wave superconducting pairing) and the critical value (both in the situations with and without p-wave pairing) for the Anderson localization phase transition will be significantly reduced, which implies an enhancement of the localization process. However, if the system is divided into two parts with one of them coupled to physical gain and the other coupled to the corresponding physical loss, the transition process will be impacted only in a very mild way. Besides, we also discuss the situations with imbalanced physical gain and loss and find that the existence of random imaginary potentials in the system will also affect the localization process while constant imaginary potentials will not.
\end{abstract}

\pacs{72.15.Rn, 11.30.Er, 03.65.Vf, 73.21.Cd}

\maketitle
\date{today}

\section{Introduction}
Open systems have gained much attention in recent years in many different research fields ranging from transport in mesoscopic systems to quantum computing. Many properties are strongly dependent on the openness of the system and the way the system interacting with the environment around it. In order to take the influences of the environment into account, the effective non-Hermitian Hamiltonian approach has been used extensively in treating open systems \cite{Mahaux, Sokolov1, Sokolov2, Dittes, Rotter, Amir}. By introducing imaginary parts to the Hamiltonian to represent the physical gain and loss of the system, one can study the open systems in an consistent way by analyzing the complex eigenvalues of the effective Hamiltonian. Among all these non-Hermitian Hamiltonians that have been explored, the $\mathcal{PT}$-symmetric ones with balanced gain and loss, which are invariant under combined parity and time-reversal operations, have drawn tremendous interest. With appropriate conditions, these $\mathcal{PT}$-symmetric Hamiltonians can have purely real energy spectra \cite{Bender1, Bender2}. The properties of $\mathcal{PT}$-symmetric Hamiltonians and the corresponding breaking of this symmetry in many different non-Hermitian systems have been extensively studied \cite{Klaiman, Sukhorukov, Ramezani, Longhi, Musslimani, Luo, Bendix, Jin, Zhu1, Wang1, Zeng1}. Experimentally, there are also many different kinds of realizations of open systems in optical \cite{Guo, Ruter, Feng, Regensburger}, mechanical \cite{Bender3}, and electrical \cite{Schindler} setups, which endow the non-Hermitian Hamiltonians more practical significance. The role of physical gain and loss in changing the properties of the open systems has been revealed and clarified in these systems.

As one of the most famous phenomena in condensed matter physics, the Anderson localization has also been extensively studied in non-Hermitian systems \citep{Amir1,Hatano1,Hatano2}. Other related properties such as the many body localization under dissipations are also explored \citep{Carmele,Levi} in open systems. Recently, the Anderson localization has been studied in disordered optical lattices systems, where it was shown that the existence of physical gain and loss could enhance the localization of light \cite{Jovic, Cortes}. Besides, the properties of $\mathcal{PT}$-symmetry in non-Hermitian Aubry-Andr\'e model have been investigated \cite{Yuce, Liang}. It is well known that the Aubry-Andr\'e model or the Aubry-Andr\'e-Harper (AAH) model will shows a phase transition from extended states to localized states (Anderson localization) when the lattice is incommensurate \cite{Harper, Aubry, Ostlund, Kohmoto}. The normal Hermitian AAH model with or without p-wave superconducting pairing presents abundant physical phenomena both in the commensurate and incommensurate situations \cite{Kraus, Lang1, Lang2, Cai, DeGottardi, Ganeshan, Satija, Barnett, Deng, Grusdt, Zhu, Zeng2, Wang}. However, the influences of physical gain and loss on the AAH model have not been explored much. In Ref. \cite{Yuce}, a $\mathcal{PT}$-symmetric Aubry-Andr\'e model is discussed and the result shows that the physical gain and loss can affect the Anderson localization. In addition, it is found that the broken $\mathcal{PT}$-symmetry can be restored via increased loss and gain in the Aubry-Andr\'e model \cite{Liang}. However, a general discussion of the non-Hermitian AAH model, especially the influences of different configurations of physical gain and loss on the Anderson localization, is still lacking. Besides, if furthermore the p-wave superconducting pairing are also introduced into these non-Hermitian AAH models, we can expect much more interesting phenomena since the corresponding Hermitian system shows a lot nontrivial properties.

In this paper, we investigate non-Hermitian Aubry-Andr\'e-Harper models with or without p-wave superconducting pairing in the presence of physical gain or loss, which are represented by imaginary potentials added to the lattice sites. We find that if the physical gain and loss are added to the even and odd sites in an alternate way, the critical region (in the case with p-wave superconducting pairing) and the critical value for the Anderson localization phase transition will be reduced significantly, which implies that the system will be easier to be localized. The Anderson localization is enhanced by the alternating physical gain and loss. However, if we divide the system into two parts, with one part coupled to certain negative (or positive) imaginary potentials while the other part coupled to the corresponding balanced positive (or negative) imaginary potentials, the localization process will just be influenced in a very mild way. Besides, we also discuss the situations with imbalanced physical gain and loss. The results show that the existence of random imaginary potentials will affect the Anderson localization transition, while if we add constant physical gain or loss in the system, the localization process will not be impacted at all. These results indicate that different configurations of physical gain and loss will have very different influences on the Anderson localization phase transition in the incommensurate Aubry-Andr\'e-Harper models.

The rest of the paper is organized as follows. In Sec. \ref{sec2}, we introduce the non-Hermitian AAH model and present the system Hamiltonian. Then we will discuss the influences of the physical gain and loss on the Anderson localization phase transition in Sec. \ref{sec3}. We will explore both AAH models with and without superconducting pairing. The last section (Sec. \ref{sec4}) is dedicated to a brief summary.

\section{Model Hamiltonian}\label{sec2}
The Hamiltonian of the one-dimensional (1D) non-Hermitian Aubry-Andr\'e-Harper model we consider in this paper can be described as
\begin{equation}\label{Eq1}
H= \sum_{j=1}^{N} V_j c_j^\dagger c_j + \sum_{j=1}^{N-1} [ -t c_{j+1}^\dagger c_j +\Delta c_{j+1}^\dagger c_j^\dagger + H.c. ]
\end{equation}
where $c_j^\dagger$ ($c_j$) is the creation (annihilation) operator at site $j$, $V_{j} = V \cos(2\pi \alpha j + \varphi_V)+i \delta_j$ is the on-site potential with $\delta_j$ being the physical gain or loss at the $j$th lattice site, $t$ is the hopping amplitude between the nearest neighboring lattice sites, and $\Delta$ is the p-wave superconducting pairing gap which is taken to be real. This one-dimensional chain has $N$ sites and the on-site potential is modulated by a cosine function with periodicity $1/\alpha$ and a phase factor $\varphi_v$. When $\alpha$ is irrational, the model system becomes quasiperiodic. If we set $\delta_j = \Delta = 0$, then the system is the famous Aubry-Andr\'e model which has been investigated for a long time. In addition, we can make this non-Hermitian system $\mathcal{PT}$-symmetric or asymmetric by assigning $\delta_j$ different values. For a 1D Hermitian AAH model with incommensurately modulated on-site potentials, the system shows an Anderson localization transition when $V$ becomes large enough [$V > 2(t+\Delta)$]. With the physical gain and loss added to the system, we may investigate the influence of the imaginary potentials on this phase transition, which is the main objective of this paper. We take $\alpha = (\sqrt{5}-1)/2$ as an example, but the results can also be generalized to other incommensurate situations. The Hamiltonian can be diagonalized by using the Bogoliubov-de Gennes (BdG) transformation \cite{Gnnes, Lieb},
\begin{equation}\label{}
  \eta_n^\dagger = \sum_{j=1}^{N} [u_{n,j} c_{n,j}^\dagger + v_{n,j} c_{n,j}],
\end{equation}
where $u_{n,j}$ and $v_{n,j}$ can be chosen to be real in the Hermitian system, but for the non-Hermitian Hamiltonian we discuss here, they are complex numbers. $n$ is the energy band index. Then the wave function of the Hamiltonian is
\begin{equation}\label{}
  | \Psi_n \rangle = \eta_n^\dagger | 0 \rangle =  \sum_{j=1}^{N} [u_{n,j} c_{n,j}^\dagger + v_{n,j} c_{n,j}] | 0 \rangle.
\end{equation}
By using the similar method in Ref. \cite{Zeng2}, we can diagonalize the Hamiltonian and get the eigenvalues and the wave function of the system.

When the lattice is incommensurate, the AAH model will show a phase transition from extended states to localized states. In order to investigate the phase transition, we can calculate the inverse participation ratio (IPR) which for a normalized wave function $\Psi_n$ is defined as $IPR=\sum_j (|u_{n,j}|^4 + |v_{n,j}|^4)$. The IPR measures the inverse of the number of occupied lattice sites and is a very useful quantity in characterizing the localization transitions of quasiperiodic systems. We further define the mean inverse participation ratio (MIPR) as $MIPR = \frac{1}{2N} \sum_{n=1}^{2N} \sum_{j=1}^{N}(|u_{n,j}|^4 + |v_{n,j}|^4)$. MIPR is close to zero when the system is in the extended states, however, it will tend to a finite value of $O(1)$ for localized states.

In the next section, we will discuss different configurations of non-Hermitian AAH model with or without balanced physical gain and loss by checking the MIPR of those systems. The Anderson localization phase transition in these AAH models shows very different behaviors, which helps us to gain more understanding of the AAH model.

\section{Numerical results and discussions}\label{sec3}
In this section, we explore the influence of physical gain and loss, which originate from the interacting processes between the system and the environment, on the Anderson localization phase transition of the AAH model. We will mainly discuss non-Hermitian AAH model with balanced and imbalanced physical gain and loss which can be altered by changing the imaginary potentials added to the lattice sites. Both the normal AAH model and the extended AAH model with p-wave superconducting pairing are investigated in these two cases.

\subsection{Case 1: Balanced physical gain and loss}
We first investigate the case with balanced physical gain and loss. To achieve the balance of physical gain and loss, we can add the positive and negative imaginary potentials to the lattice sites in an alternate way, namely we have
\begin{equation}
\delta_j = \begin{cases}
-i \delta,& \text{j odd}\\
i \delta.& \text{j even}
\end{cases}
\end{equation}
In this situation, there are both physical gain and loss in the AAH model and they are added in an alternate way to the odd and even lattice sites. The physical gain and loss thus are balanced in the system. When the site number is even, the Hamiltonian can be $\mathcal{PT}$-symmetric if we set $\varphi_V=-\pi \alpha (N+1)$ for the incommensurate lattice, as suggested in Ref. \cite{Yuce}. It is well known that some non-Hermitian Hamiltonians can have totally real energy spectra if they are $\mathcal{PT}$-symmetric. Here we just focus on the properties of the Anderson localization and don't pay too much attention to the $\mathcal{PT}$-symmetry of the system, so we set $\varphi_V=0$ throughout this paper.

\begin{figure}[!ht]
\centering
\subfigure[$\Delta=0$]{
\label{fig1a}
\includegraphics[width=3.0in]{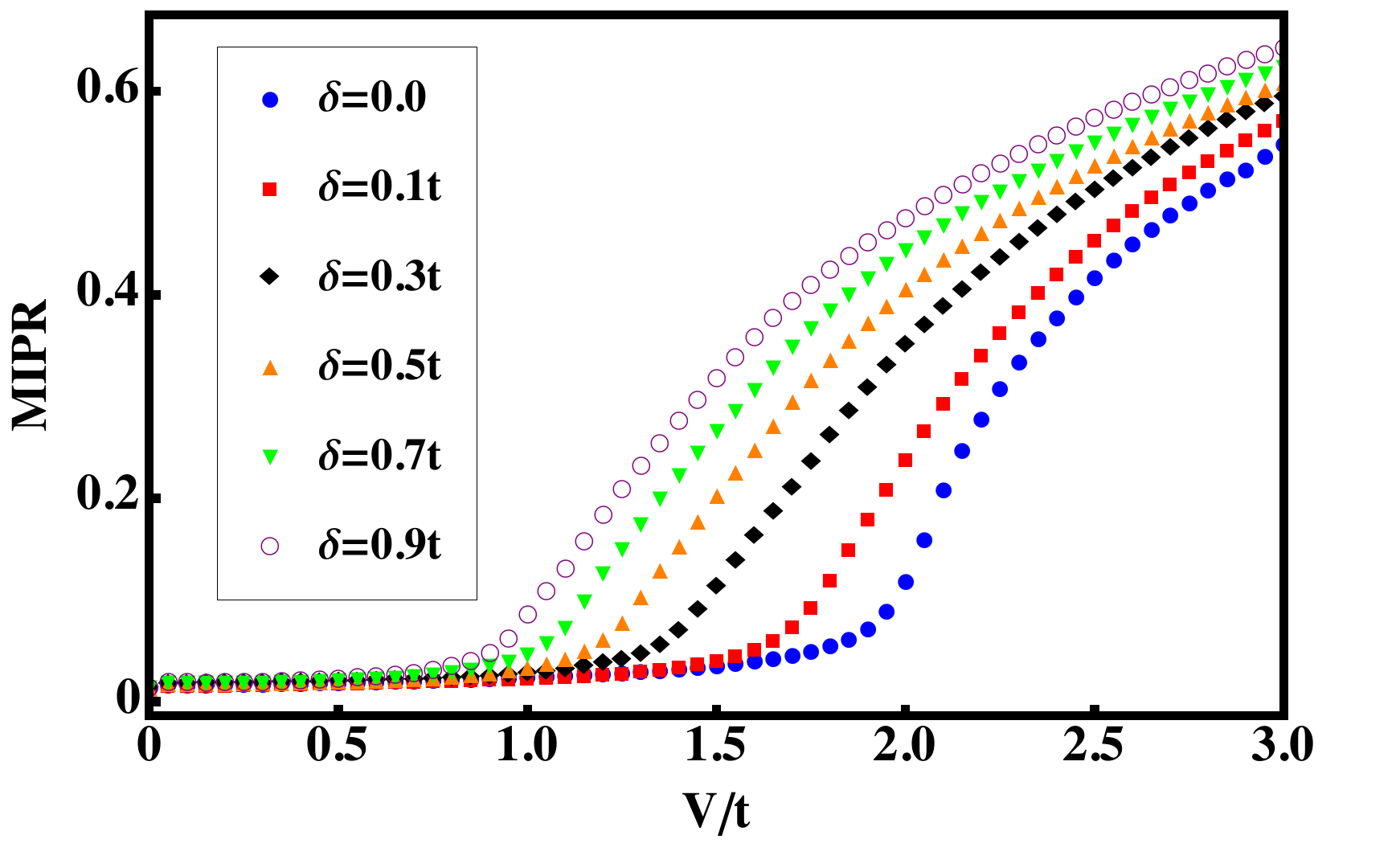}}
\subfigure[$\Delta=0.2t$]{
\label{fig1b}
\includegraphics[width=3.0in]{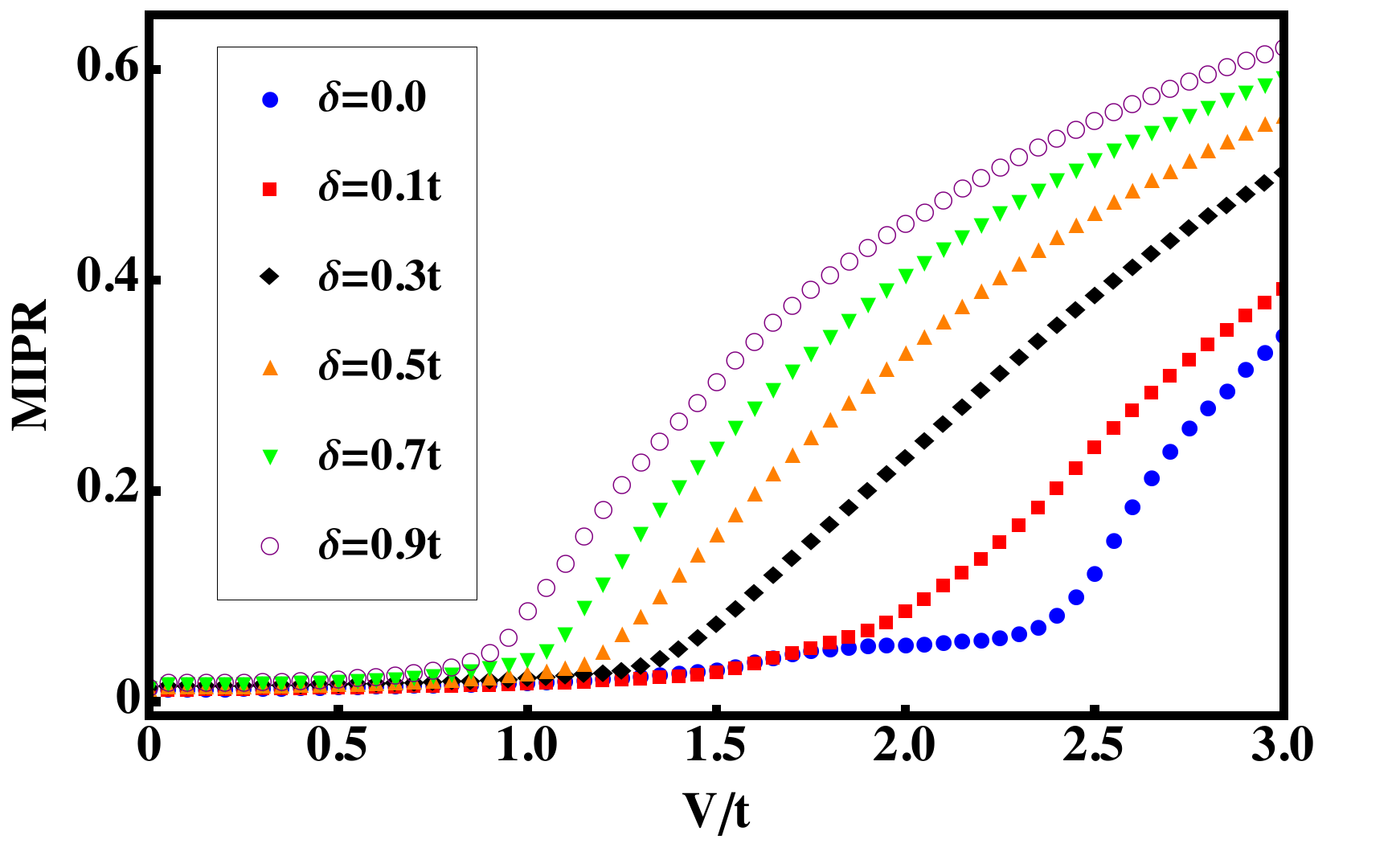}}
\caption{(Color online) MIPR of a non-Hermitian AAH model with negative and positive imaginary potentials added alternately to the lattice sites. (a) MIPR vs V when $\Delta=0$; (b) MIPR vs V when $\Delta=0.2t$. The lattice site number is $N=100$ and $\varphi_V=0$.}
\label{fig1}
\end{figure}

Now we investigate the influence of the physical gain and loss introduced by the alternate imaginary potentials on the Anderson localization phase transition of this incommensurate system. The MIPR we introduced earlier can be used to identify the phase transition. It is well known that when there is no superconducting (SC) pairing, the normal AAH model shows a phase transition from extended states to localized states directly at some critical $V$ value. However, if SC pairing is introduced, there will be a critical region before the system turns into the localized states \cite{Zeng2, Wang}. Fig. \ref{fig1} shows the MIPR of the system as a function of $V$ in situations with and without SC pairing terms. When $\delta=0$, the system is Hermitian. The critical value for the phase transition is $V = 2t$ when $\Delta=0$ and $V = 2(t+\Delta)$ when $\Delta \neq 0$, as shown by the blue dots in Fig. \ref{fig1a} and \ref{fig1b}. These results are consistent with the conclusions from previous study \cite{Cai, Zeng2}. Besides, the plateau in the MIPR of the AAH model with SC pairing represents the critical region during the phase transition process, see Fig. \ref{fig1b}. As the imaginary potential increases, this critical region reduces gradually and disappear in the end. The critical value for the phase transition $V_c$ decreases both in the $\Delta=0$ and $\Delta \neq 0$ case. Though the lattice number here is chosen to be $N=100$, the results are almost the same as those in larger systems. Thus we can conclude that by adding the physical gain and loss alternately to the lattice sites, the incommensurate AAH model will be easier to be localized. The localization is enhanced by the alternate physical gain and loss. The reason behind this is that with $+i\delta$ and $-i\delta$ alternatively added to the lattice sites, the wave function for an electron being at certain lattice site will acquire a phase factor which will lead to an enhancement in the probability for the electron being at that site as the system evolves in time, while the wave functions for its neighboring sites acquire a phase factor which will lead to a reduction in the probability for being at these sites. So the electrons in this system will tend to stay at some lattice sites with much higher probability as the system evolves, which makes the system easier to be localized.

\begin{figure}[!ht]
\centering
\subfigure[$\Delta=0$]{
\label{fig2a}
\includegraphics[width=3.0in]{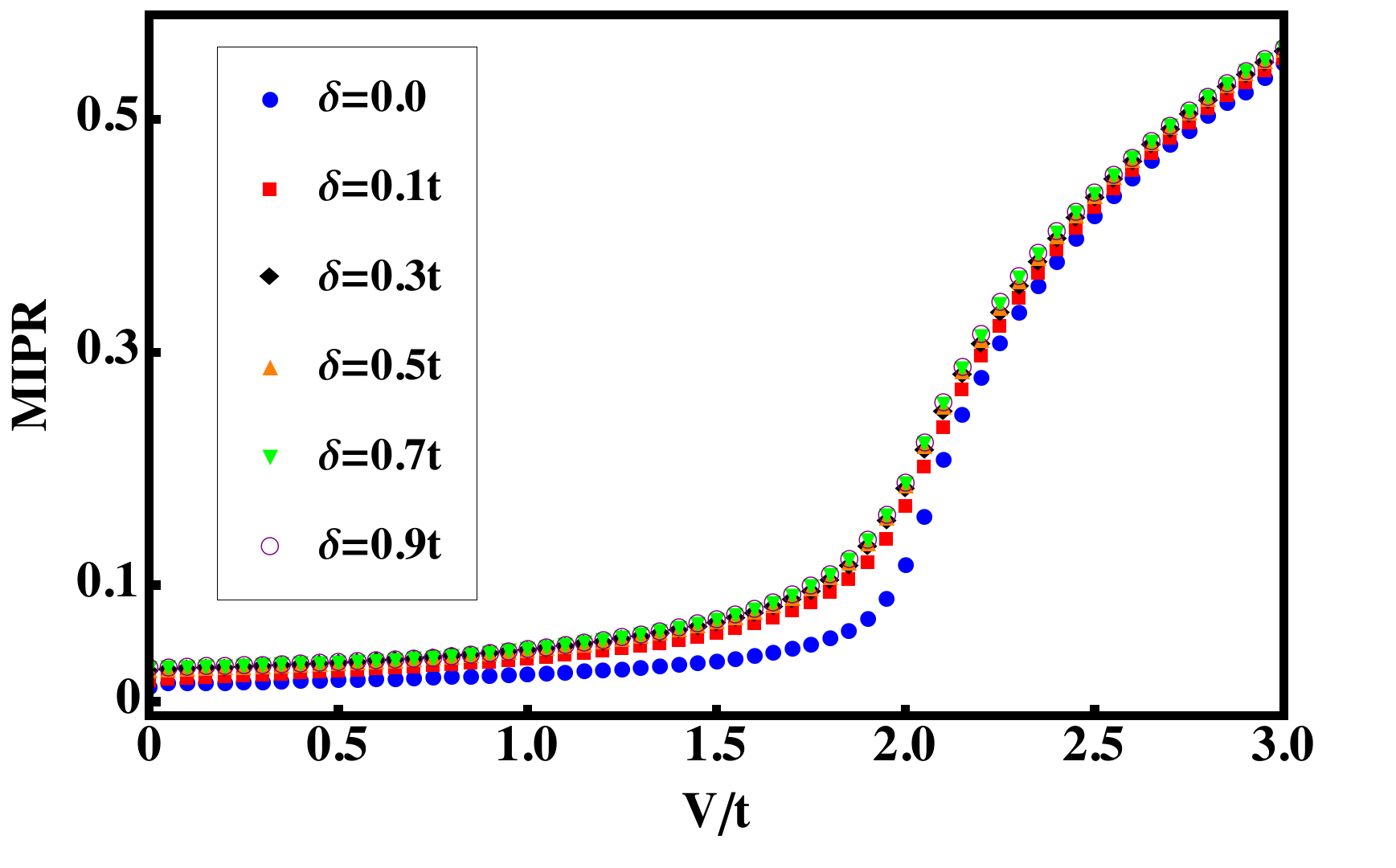}}
\subfigure[$\Delta=0.2t$]{
\label{fig2b}
\includegraphics[width=3.0in]{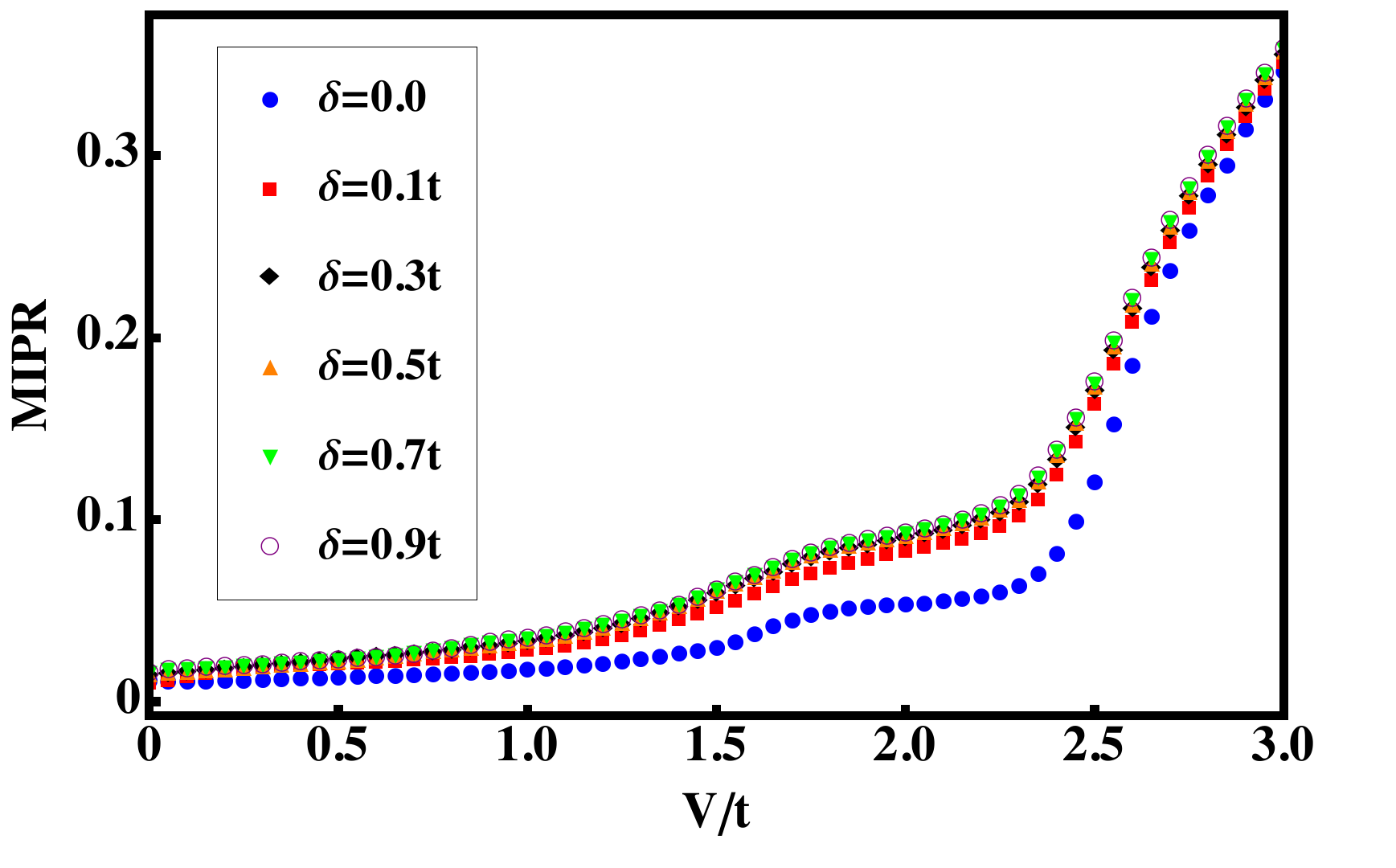}}
\caption{(Color online) Here the whole system is divided into two parts with the same number of lattice sites. The imaginary potential $-i \delta$ is added to the lattice sites in one part of the system while $+i \delta$ is added to the lattice sites in the other part. (a) MIPR vs V when $\Delta=0$; (b) MIPR vs V when $\Delta=0.2t$. The lattice site number is $N=100$ and $\varphi_V=0$.}
\label{fig2}
\end{figure}

On the other hand, we can also achieve the balance of physical gain and loss in another way. We divide the whole system into two parts each with $N/2$ lattice sites. Then the negative imaginary potential $-i \delta$ is added to the sites in one part while the positive imaginary potential $+i \delta$ is coupled to the sites in the other part, so we have
\begin{equation}
\delta_j = \begin{cases}
-i \delta,& \text{$j \leq N/2$}\\
i \delta.& \text{$j > N/2$}
\end{cases}
\end{equation}
Thus the gain and loss are also balanced. The MIPR of this kind of model is presented in Fig. \ref{fig2}. It is interesting to notice that even though the MIPR of the system is also changed, but only in a very mild way. The influence of the physical gain and loss on the localization process in this situation is much weaker than that in the case with alternate physical gain and loss. Since now the imaginary potentials added to the lattice site in either part of the bipartite system are the same, the relative probability for an electron stay at the sites of the same part will be the same, except those sites near the boundary of the two different parts. So by applying the physical gain and loss in such a way, there will be very little impact on the Anderson localization process of the system.

\subsection{Case 2: Imbalanced physical gain and loss}
If we add only positive or negative imaginary potentials to every lattice site of the model, for example placing the system into an environment with only physical gain or physical loss, the eigenvalues of the system will always be complex since we have a $\pm\delta$ term in the Hamiltonian and the imaginary of the energy spectra will always be $\pm\delta$.

\begin{figure}[!ht]
\centering
\subfigure[$\Delta=0.0$]{
\label{fig3a}
\includegraphics[width=3.0in]{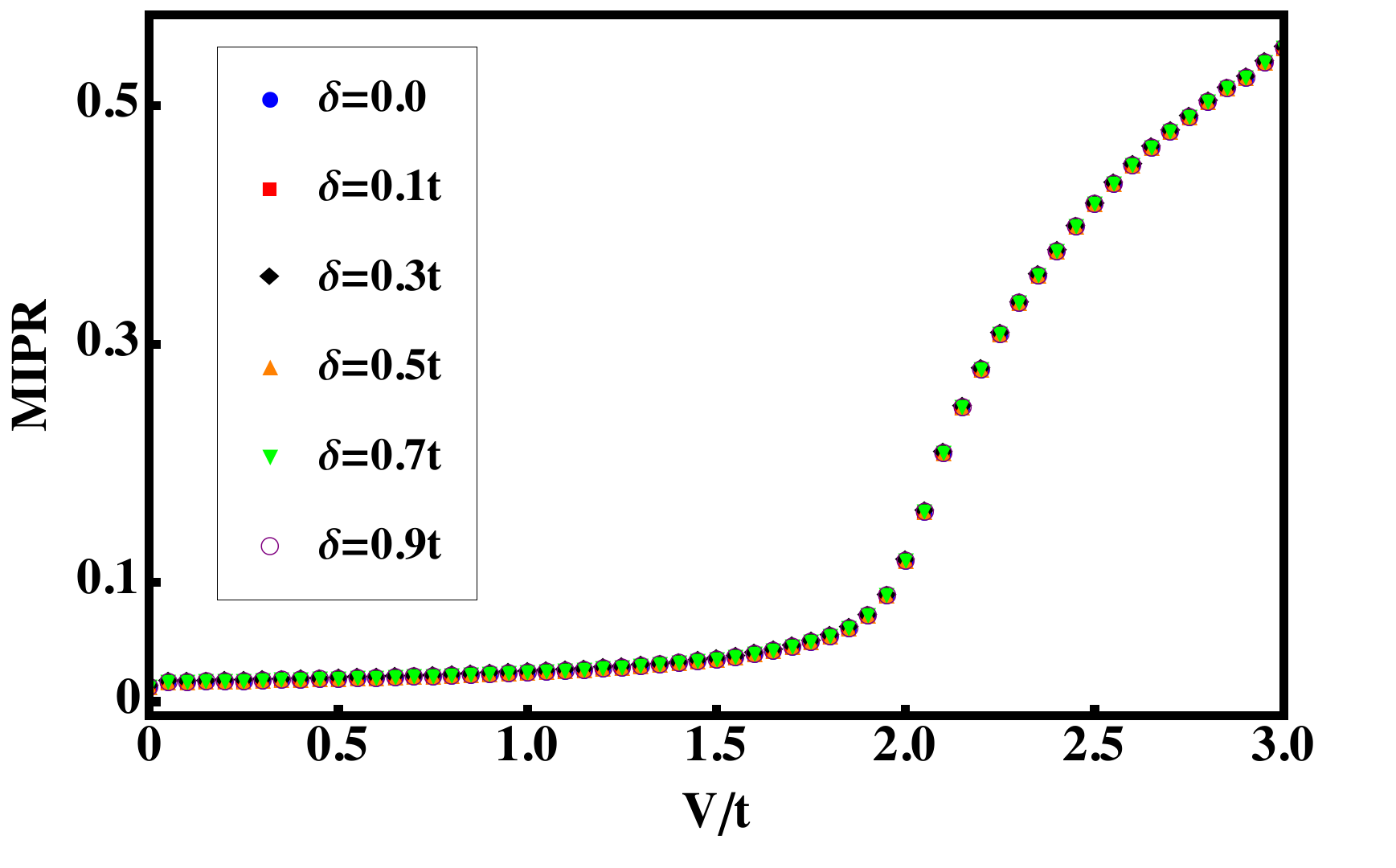}}
\subfigure[$\Delta=0.2t$]{
\label{fig3b}
\includegraphics[width=3.0in]{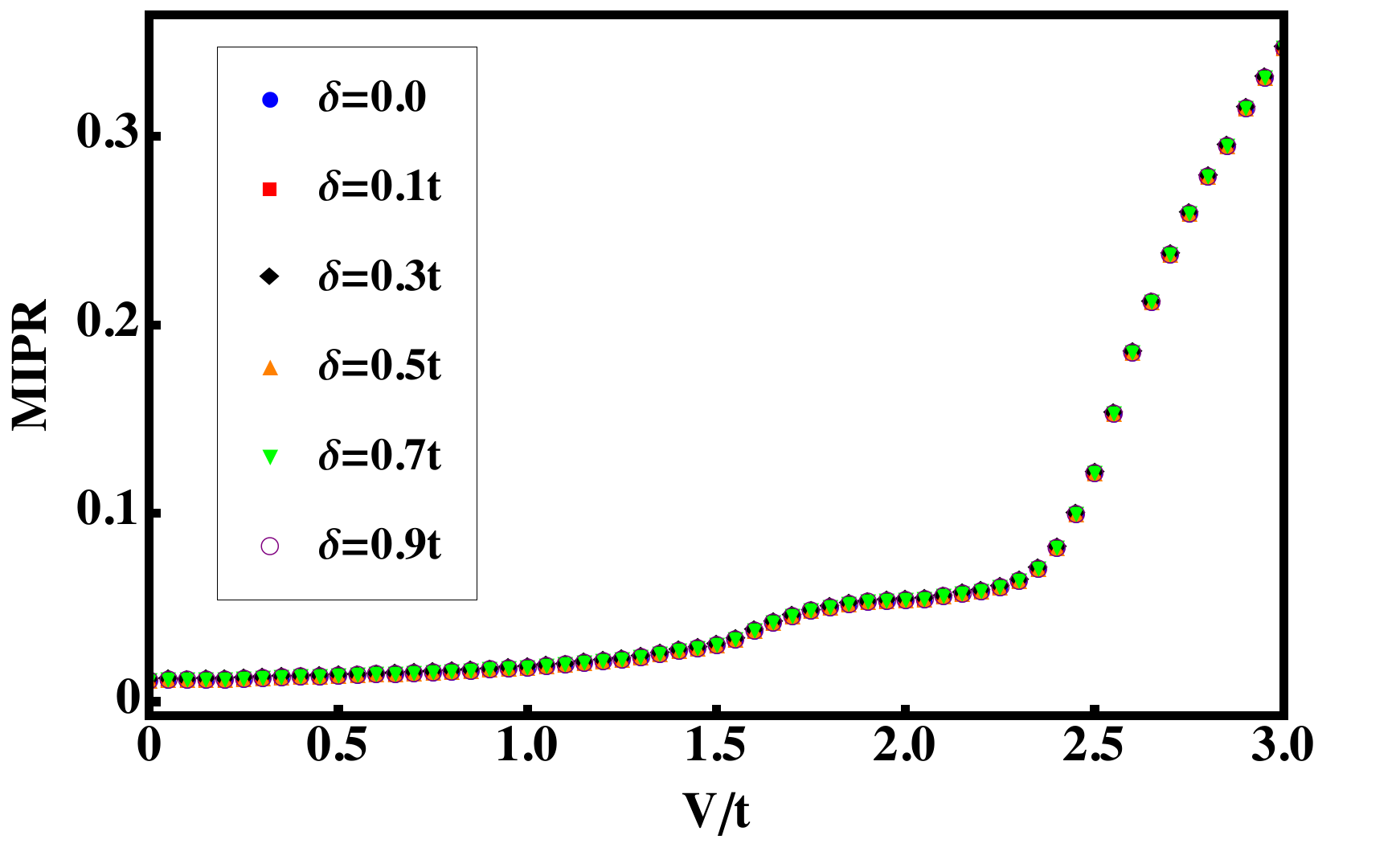}}
\caption{(Color online) MIPR as a function of $V$ for a AAH model with constant physical gain or loss imaginary potential $-i\delta$ added to all lattice sites. Similar results can also be applied to the model with imaginary potential $+i\delta$. Here we also set $N=100$ and $\varphi_V = 0$.}\label{fig3}
\end{figure}

Then we check the influence of constant physical gain or loss on the Anderson localization phase transition. We also calculate the MIPR of the system and explore its variation due to the imaginary potential $-i\delta$. The results are given in Fig. \ref{fig3}. It is clear from Fig. \ref{fig3a} and \ref{fig3b} that if the imaginary potentials are the same on all lattice sites, the MIPR are identical to the Hermitian case (i.e. $\delta=0$), which means that the critical value as well as the critical region can not be changed at all. The same conclusion can be applied to the non-Hermitian AAH model with imaginary potential $+i\delta$ added to all the lattice sites. So the localization process of the AAH model will not be affected in an environment with constant physical gain or loss. This is understandable since with the same physical gain or loss at every site, the wave functions acquire a same phase factor for all the sites, thus the relative probabilities for electrons being at certain lattice site remain unchanged even when the imaginary potential becomes stronger.

\begin{figure}[!ht]
\centering
\subfigure[$\Delta=0.0$]{
\label{fig4a}
\includegraphics[width=3.0in]{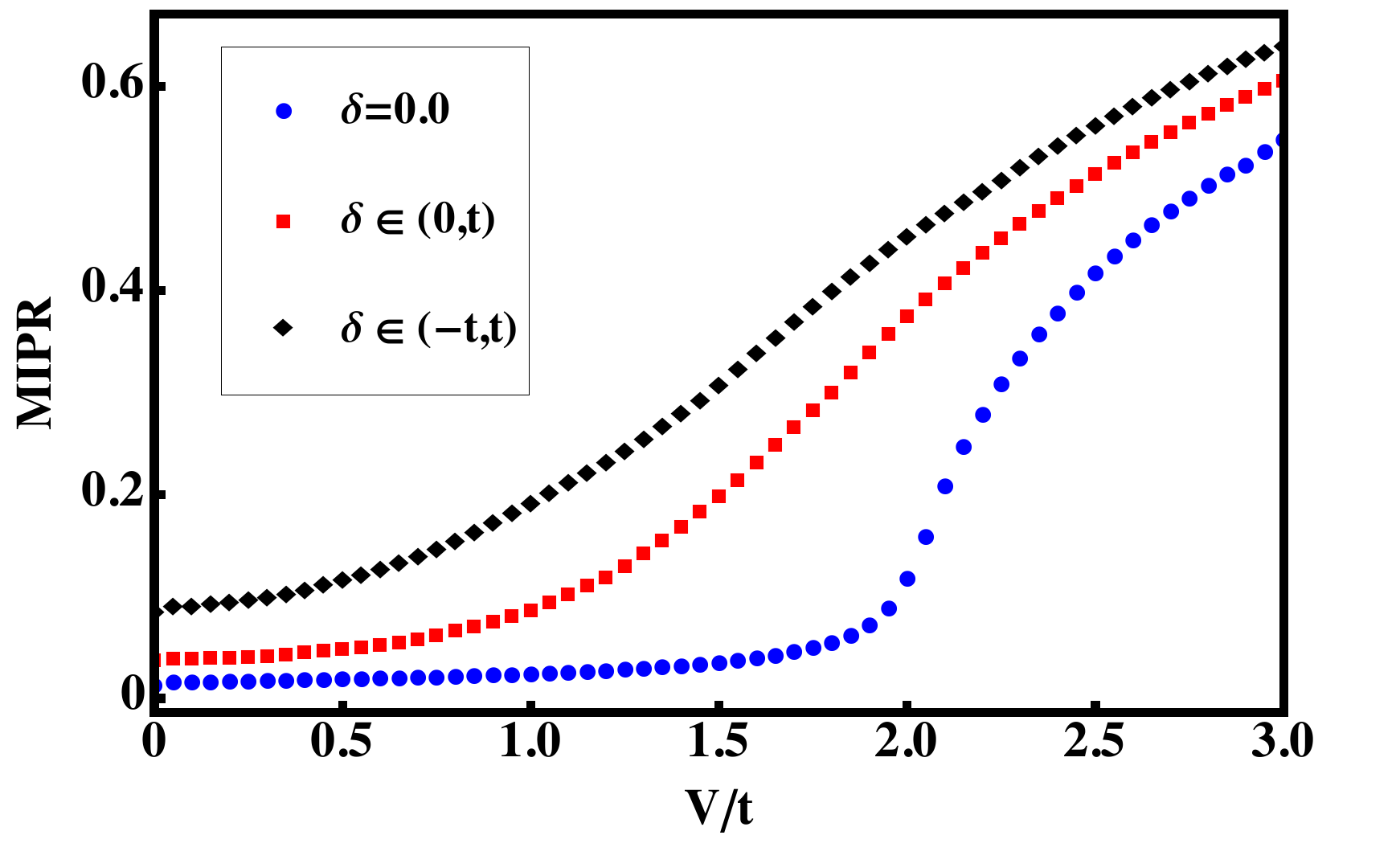}}
\subfigure[$\Delta=0.2t$]{
\label{fig4b}
\includegraphics[width=3.0in]{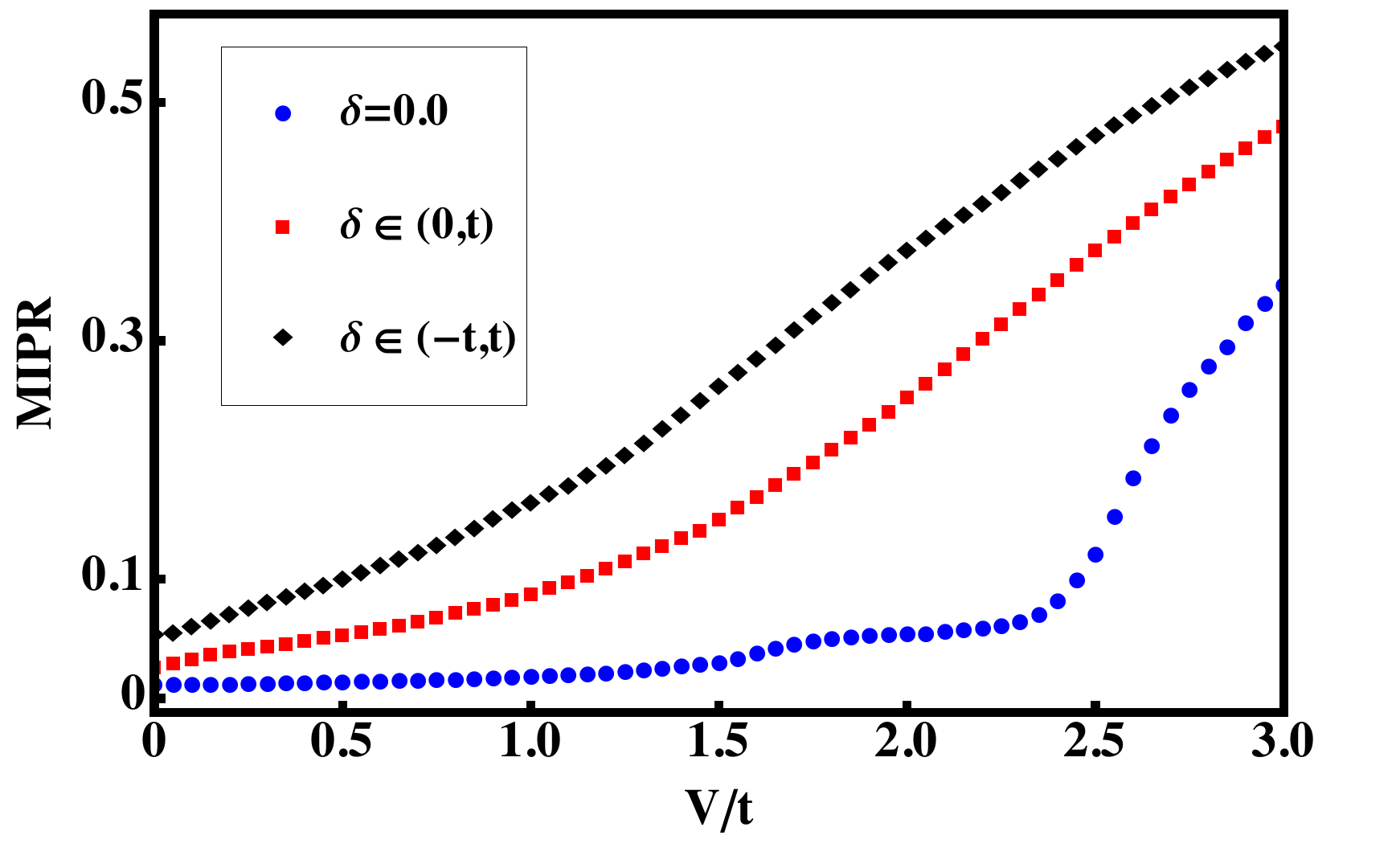}}
\caption{(Color online) MIPR as a function of $V$ for a AAH model with zero or random imaginary potentials added to the lattice sites. The blue dots correspond to the situation without physical gain and loss ($\delta_j=0$). For the situations with random imaginary potentials, the range of $\delta_j$ is $(0,t)$ (red dots) and $(-t,t)$ (black dots), respectively. The number of lattice sites is $N=100$.}\label{fig4}
\end{figure}

Next, we turn to another case with imbalanced physical gain and loss. We set the imaginary potentials of the AAH model randomly and calculate the MIPR of the system. We can expect that the localization process will also be influenced in this situation since the relative probability of the electrons being at some lattice sites will be quite different from other sites due to the randomness of the physical gain and loss. In Fig. \ref{fig4}, we present the results of such AAH models with or without p-wave superconducting pairing. When we compare it with the MIPR when $\delta=0$ (blue points in the figure), it is obvious that the system with random physical gain and loss is easier to be localized. It is also noteworthy that if we set $\delta_j$ to be a random value between $(-t,t)$, the MIPR will be larger than that when we set $\delta_j$ to be a random value between $(0,t)$, as shown by the black and red dots in Fig. \ref{fig4a} and \ref{fig4b}. So with stronger randomness in the imaginary potentials, the system is easier to be localized.

\section{Summary}\label{sec4}
In this paper, we have investigated the non-Hermitian Aubry-Andr\'e-Harper model with physical gain and loss which are represented by the imaginary potentials added to the lattice sites. We mainly focused on two different kinds of non-Hermitian AAH model with or without p-wave superconducting pairing. In the situation with balanced physical gain and loss, we find that the Anderson localization of the incommensurate AAH model can be impacted appreciably when the physical gain and loss are added to the even and odd lattice sites in an alternate way, the critical region and the critical value for the transition are reduced due to the physical gain and loss, which means that the Anderson localization is enhanced. However, if the balanced physical gain and loss is applied to the system in another way, where the system is divided into two parts with one part coupled to negative (positive) imaginary potentials and the other part coupled to the corresponding positive (negative) imaginary potentials, the phase transition will only be influenced in a very mild way. In another case with imbalanced physical gain and loss, we find that the phase transition is not affected when all the lattice sites are added to the same imaginary potential while the existence of random physical gain and loss will influence the localization process. All these results indicate that different configurations of the physical gain and loss will have very different impact on the Anderson localization process in the AAH model. We hope that the conclusions obtained in the present work will stimulate more interest and research on the non-Hermitian AAH model and the Anderson localization under the influences of environment. \\

\section*{Acknowledgments}
This work has been supported by the NSFC under Grant No. 11274195 and the National Basic Research Program of China (973 Program) Grants No. 2011CB606405 and No. 2013CB922000. S. C. is supported by NSFC under Grants No. 11425419, No. 11374354 and No. 11174360, and the Strategic Priority Research Program (B) of the Chinese Academy of Sciences (No. XDB07020000).

\end{document}